\documentclass[prd,aps,preprintnumbers, showpacs, nofootinbib,superscriptaddress,notitlepage]{revtex4-1}

\usepackage{latexsym,bm,amsmath,amssymb,amsfonts}
\usepackage{graphicx}
\usepackage{amsmath,amsthm,amssymb}

\newcommand{\beq}{\begin{eqnarray}}
\newcommand{\eeq}{\end{eqnarray}}

\newcommand{\nn}{\nonumber \\}

\begin{document}

\preprint{YITP-17-56}
\preprint{LU TP 17-25}

\title{Accessing the gluon Wigner distribution in ultraperipheral $pA$ collisions}

\author{Yoshikazu Hagiwara} 
\affiliation{Department of Physics, Kyoto University, Kyoto 606-8502, Japan}
\email{hagiyoshihep@gmail.com}

\author{Yoshitaka Hatta} 
\affiliation{Yukawa Institute for Theoretical Physics, Kyoto University, Kyoto 606-8502, Japan}
\email{hatta@yukawa.kyoto-u.ac.jp}

\author{Roman Pasechnik}
\affiliation{Department of Astronomy and Theoretical Physics, Lund University, SE-223 62 Lund, Sweden}
\email{Roman.Pasechnik@thep.lu.se}

\author{Marek Tasevsky}
\affiliation{Institute of Physics, Academy of Sciences of the Czech Republic, 18221 Prague 8, Czech Republic}
\email{Marek.Tasevsky@cern.ch}

\author{Oleg Teryaev}
\affiliation{Joint Institute for Nuclear Research, 141980 Dubna, Russia}
\email{teryaev@theor.jinr.ru}

\begin{abstract}
We propose to constrain the gluon Wigner distribution in the nucleon by studying the exclusive diffractive dijet production process in ultraperipheral 
proton-nucleus collisions (UPCs) at RHIC and the LHC. Compared to the previous  proposal  in Ref.~[Y. Hatta, B. W. Xiao, and F. Yuan, Phys. Rev. Lett. 
116, 202301 (2016).] to study the same observable in lepton-nucleon scattering, the use of UPCs has a few advantages: not only is the cross section larger, 
but the extraction of the Wigner distribution from the data also becomes simpler, including its elliptic angular dependence. We compute the corresponding 
cross section and evaluate the coefficients using models which include the gluon saturation effects. A potential for the measurements of the Wigner distribution 
at current and future experimental facilities is also discussed.
\end{abstract}

\pacs{12.38.-t,12.38.Lg,12.39.St,13.60.-r,13.85.-t}

\maketitle

\section{Introduction}
\label{Sec:Intro}

The so-called Wigner distribution is known to provide maximally detailed information on quantum systems describing the distribution of particles in phase space. 
In the case of hadron structure, the QCD Wigner distribution~\cite{Ji:2003ak,Belitsky:2003nz,Lorce:2011kd}, or its Fourier transform, the Generalized Transverse Momentum 
Dependent Distribution (GTMD) \cite{Meissner:2009ww,Hatta:2011ku,Lorce:2013pza,Echevarria:2016mrc}, provides multidimensional partonic imaging of the nucleon (for a detailed 
review on this topic, see e.g.~Refs.~\cite{Boer:2011fh,Accardi:2012qut}). It gives the most comprehensive description of hadron structure (parton tomography) and, 
as it is not calculable in perturbative QCD, the question of its measurement naturally arises.

The measurement of various nonperturbative ingredients of QCD factorization (``partonometry'') is in general a challenging problem. While spin-averaged and 
spin-dependent parton distributions can be studied in (inclusive) Deep Inelastic Scattering (DIS), the studies of the Transverse Momentum Dependent Distributions (TMDs) 
rely mostly on  semi-inclusive DIS (SIDIS), and the Generalized Parton Distributions (GPDs) are extracted from the data on exclusive processes, mostly Deeply Virtual Compton 
Scattering (DVCS). However, these processes are sensitive to either the transverse momentum $\vec{q}_\perp$ or impact parameter $\vec{b}_\perp$ of partons, whereas 
the Wigner distribution  $W(x,\vec{q}_\perp,\vec{b}_\perp)$ depends on both\footnote{Here, $x$ denotes the longitudinal momentum fraction. Throughout this paper, 
we suppress the dependence on the skewness parameter $\xi$. In the small-$x$ region which we are interested in, effectively $\xi\approx x$.}. Is there 
a way to phenomenologically access such detailed information on parton tomography in the nucleon?

Recently, new observables to measure gluon GTMDs in the small-$x$ region in exclusive diffractive dijet production at an electron-ion collider (EIC) have been proposed 
in Ref.~\cite{Hatta:2016dxp} (see also a related work \cite{Altinoluk:2015dpi})\footnote{More recently, a method to access the quark GTMDs for generic values of $x$ 
in the exclusive double Drell-Yan process has been proposed \cite{Bhattacharya:2017bvs}.}. In particular, it was understood that the gluon GTMD distribution at small-$x$ 
can be considered as a Fourier transform of an impact parameter dependent forward dipole amplitude (or dipole $S$-matrix), which  provides access to 
the gluon saturation effects at small-$x$ (see e.g. Ref.~\cite{Gelis:2010nm}). Moreover, the process is also sensitive to the characteristic azimuthal angular correlation 
between $\vec{q}_\perp$ and  $\vec{b}_\perp$ governed by the ``elliptic'' gluon Wigner distribution \cite{Hatta:2016dxp,Hagiwara:2016kam,Zhou:2016rnt}. The actual 
measurement of the proposed  observables in lepton-nucleon scattering is challenging, as it requires reconstruction of full dijet kinematics vetoing any other hadronic activity
in order to reduce the backgrounds associated with the Pomeron and photon breakup. In addition, it is mandatory to detect the forward proton to ensure exclusivity of the diffractive 
process. While these experimental challenges are likely to be overcome at the planned EIC, the extraction of the GTMD is further complicated by the fact that the cross section 
is not directly proportional to the GTMD, but is given by its convolution integral which is difficult to invert. It is thus worthwhile to look for other processes in which  the latter 
problem becomes  simpler.  The vast experimental data on hadronic and nuclear collisions are now emerging from the LHC, and it would be very desirable to exploit them for 
GTMD studies. We will show below that diffractive dijet production in ultraperipheral $pA$ collisions (UPCs) at the LHC and at the RHIC is a particularly important example that 
provides an essential means for such studies. 

In UPCs the relativistic colliding systems (such as nucleons and nuclei) pass each other at large transverse distances without interacting hadronically, only
electromagnetically through the emission of quasi-real Weisz$\ddot{{\rm a}}$cker-Williams (WW) photons \cite{vonWeizsacker:1934nji,Williams:1934ad}. 
The effective WW photon flux of a charged particle is scaled as the square of its charge and thus is noticeably enhanced for heavy ions making UPCs in $pA$ 
more advantageous compared to those in $pp$. Besides, the WW spectrum is rather broad with the maximal photon energy in the target rest frame scaling linearly
with the nuclear Lorentz factor. In addition, UPCs in $pA$ provide good experimental opportunities for  studies of exclusive diffractive observables 
by detecting the intact  protons and possibly also ions using the LHC forward proton spectrometers (such as Roman pots in TOTEM~\cite{Anelli:2008zza}, 
CT-PPS~\cite{CT-PPS} at the CMS side or ALFA~\cite{ALFA} and AFP~\cite{AFP,Tasevsky:2015xya} at ATLAS side). Together with measurements of the diffractive 
dijet system, the latter would enable full kinematic reconstruction by identifying the momentum transfers from the proton and the ion separately\footnote{It should be, 
however, noted that detailed feasibility studies for the double-tagging in $p$+Pb runs are still to be performed.}. Due to a large relative distance between the scattering 
particles, the measurements of UPCs in $pA$ can be performed with no significant event pileup and with an efficient subtraction of nonexclusive diffractive backgrounds 
(for more details, see e.g.~Refs.~\cite{Bertulani:2005ru,Salgado:2011wc}).
\begin{figure*}[!t]
 \centerline{\includegraphics[width=0.4\textwidth]{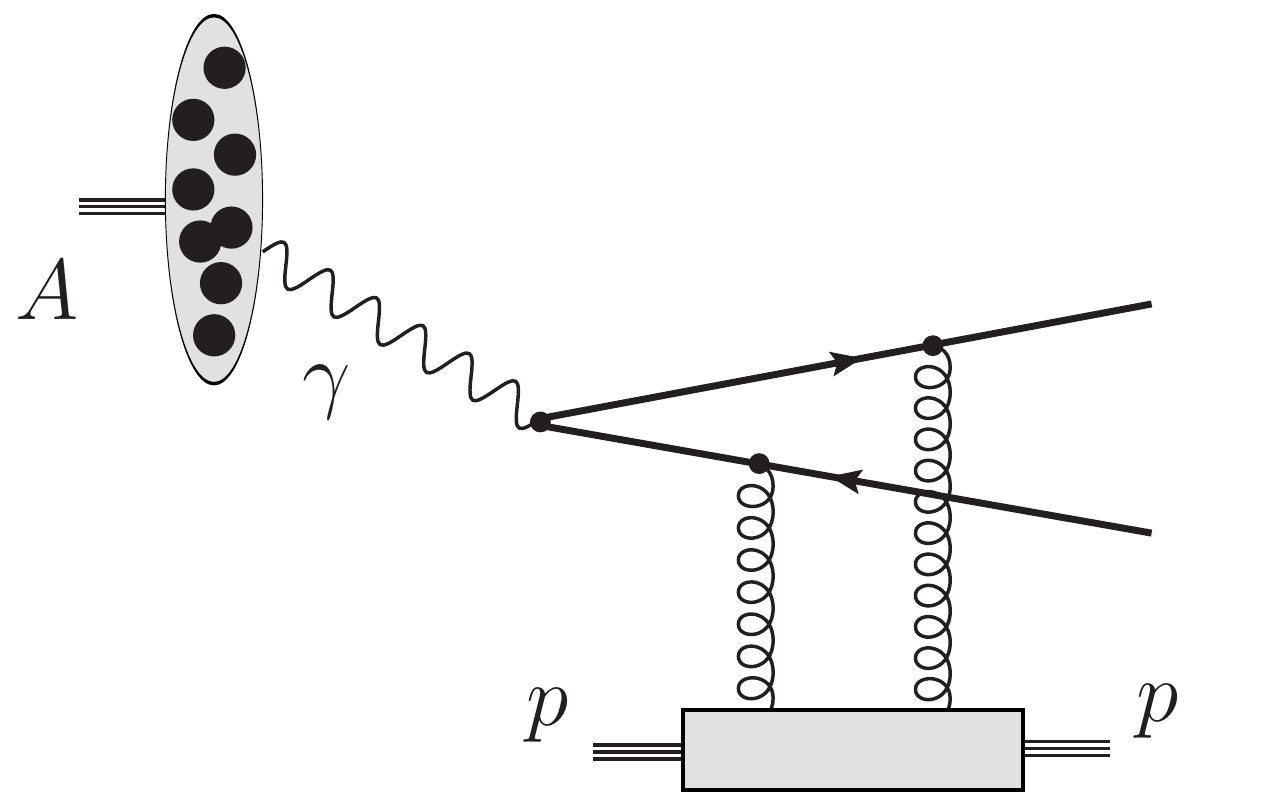}}
   \caption{
   \small A typical leading-order contribution to the exclusive diffractive dijet production in $pA$ UPCs.}
 \label{fig:dijet-UPC}
\end{figure*}

The exclusive diffractive dijet photoproduction in UPCs to the next-to-leading order at the LHC was studied recently in Ref.~\cite{Guzey:2016tek}.
Compared to typical DIS kinematics at HERA, it was understood that this process exhibits an enhanced sensitivity to small momentum fractions of the Pomeron 
exchange and a significant extension in the invariant mass (or c.m.~energy) of the photon-proton system. In the dipole picture, this process to the leading order 
can be viewed as a fluctuation of the projectile photon into its lowest Fock state, a $q\bar q$ dipole with intrinsic separation $\vec{r}_\perp$ that scatters off 
the gluon field in the nucleon target at impact parameter $\vec{b}_\perp$ by means of a color-singlet di-gluon exchange at small-$x$. Such a process is illustrated 
by a representative leading-order diagram in Fig.~\ref{fig:dijet-UPC}. 

Due to color screening, at vanishing dipole sizes $|\vec{r}_\perp|\ll R_{\rm had}$ compared to the typical hadronic scale $R_{\rm had}$ the partial (elastic) dipole amplitude 
vanishes quadratically as $\propto |\vec{r}_\perp|^2$ as the essence of the color transparency. In the opposite limit of large $|\vec{r}_\perp|\sim R_{\rm had}$, the dipole 
amplitude levels off at a certain momentum scale $Q_s^2$ known as the saturation scale that is generally dependent on the momentum fraction of the diffractive exchange 
or on $\gamma p$ c.m.~energy. While the elliptic component of the Wigner distribution characterizes the dependence of the corresponding partial dipole amplitude on 
the azimuthal angle $\phi$, the $\phi$-independent part of the Wigner distribution determines the unintegrated gluon density in the target nucleon, and at the same 
time both components are strongly sensitive to the saturation dynamics \cite{Hagiwara:2016kam,Iancu:2017fzn}. 

In this paper, we explore the differential observables of exclusive diffractive dijet production in $pA$ UPCs, $pA \to p+jj+A$  (and possibly also in 
$AA \to A+jj+A$ \cite{Kotko:2017oxg}),  and show that both components of the gluon Wigner distribution (and thus the corresponding GTMD) may be efficiently 
extracted from such data. Due to the many advantages of $pA$ UPCs described above, the considered process offers plausible opportunities for a measurement of 
the elliptic component that is only a few percent effect. In making the predictions for such a measurement, we have employed the McLerran-Venugopalan (MV) model 
\cite{McLerran:1993ka} for the gluon distribution in the target accounting for an inhomogeneity in the transverse plane following Ref.~\cite{Iancu:2017fzn}, as well as 
the dipole $S$-matrix at small-$x$ obtained by a solution of the Balitsky-Kovchegov (BK) equation \cite{Balitsky:1995ub,Kovchegov:1999yj} in the impact parameter space 
following Ref.~\cite{Hagiwara:2016kam}. 

The paper is organized as follows. In Section~\ref{Sec:diff-jj} we discuss the formulation of exclusive diffractive dijet production in $pA$ UPCs 
in terms of the dipole $S$-matrix. Section~\ref{Sec:model-calc} is devoted to a discussion of numerical results for the diffractive dijet observables 
based upon the predictions for the dipole $S$-matrix that employ the Balitsky-Kovchegov equation (for the proton target) and the McLerran-Venugopalan model 
(for a large nuclear target). Further prospects for experimental measurements of the Wigner distribution are given in Section~\ref{Sec:Exp}. Finally,
some concluding remarks are summarized in Section~\ref{Sec:Conclusions}.

\section{Exclusive diffractive dijet production in $pA$ UPCs}
\label{Sec:diff-jj}

The dipole gluon Wigner distribution is defined as
\beq
xW(x,\vec{q}_\perp,\vec{b}_\perp) &=& \frac{2}{P^+(2\pi)^3 }\int dz^+ 
d^2\vec{z}_\perp \int \frac{d^2\vec{\Delta}_\perp}{(2\pi)^2}  
e^{i\vec{q}_\perp \cdot \vec{z}_\perp  -ixP^-z^+}  \nn 
&& \times \left\langle P+\tfrac{\vec{\Delta}_\perp}{2} \left|  {\rm Tr}\, 
\left[U_+F_a^{+i}\left(\vec{b}_\perp+\tfrac{z}{2}\right) U_ - 
F_a^{+i}\left(\vec{b}_\perp-\tfrac{z}{2}\right)\right] \right|P -
\tfrac{\vec{\Delta}_\perp}{2}\right\rangle \,, \label{wigner}
\eeq
where $|P\rangle$ is the proton state and  $U_{\pm}$ is the staple-shaped  Wilson line which goes to light-cone infinity $z^+=\pm \infty$ and 
comes back. The GTMD distribution $xW(x,\vec{q}_\perp,\vec{\Delta}_\perp)$ is then given by the Fourier transform $\vec{b}_\perp \to \vec{\Delta}_\perp$. 
The key observation of Ref.~\cite{Hatta:2016dxp} is that the gluon GTMD distribution at small-$x$ is proportional to 
the Fourier transform of the dipole $S$-matrix
\beq
xW(x,\vec{q}_\perp,\vec{\Delta}_\perp) \approx \frac{2N_c}{\alpha_s} 
\left(q_\perp^2 -\frac{\Delta_\perp^2}{4}\right)
S_Y(\vec{q}_\perp,\vec{\Delta}_\perp)\,, \label{st}
\eeq
where $Y\equiv \ln 1/x$ is the rapidity and 
\beq 
S_Y(\vec{q}_\perp,\vec{\Delta}_\perp) = \int \frac{d^2\vec{r}_\perp d^2\vec{b}_\perp}{(2\pi)^4}
e^{i\vec{\Delta}_\perp \cdot \vec{b}_\perp + i\vec{q}_\perp \cdot \vec{r}_\perp}
\left\langle \frac{1}{N_c}{\rm Tr}\, U\left(\vec{b}_\perp+\frac{\vec{r}_\perp}{2}\right)
U^\dagger\left(\vec{b}_\perp -\frac{\vec{r}_\perp}{2}\right)\right\rangle_Y\,,
\eeq 
in terms of the lightlike Wilson line $U$ in the fundamental representation and the number of QCD colors $N_c=3$. Eq.~(\ref{st}) shows that the measurement 
of the GTMD distribution boils down to that of the dipole $S$-matrix. In order to be sensitive to {\it both} $\vec{q}_\perp$ and $\vec{\Delta}_\perp$, it has been 
suggested in Ref.~\cite{Hatta:2016dxp} to measure exclusive diffractive dijet production in lepton-nucleon scattering in which the proton scatters elastically 
with momentum transfer $\vec{\Delta}_\perp$ and the virtual photon splits into a $q\bar{q}$ pair (dipole) and then hadronizes into a dijet in the forward region 
with transverse momenta $\vec{k}_{1\perp}$ and $\vec{k}_{2\perp}$ such that $\vec{k}_{1\perp}+\vec{k}_{2\perp}=-\vec{\Delta}_\perp$. By measuring 
the differential cross section as a function of the relative transverse momentum of the dijet $\vec{P}_\perp = \frac{1}{2}(\vec{k}_{2\perp}-\vec{k}_{1\perp})$ 
at fixed $\vec{\Delta}_\perp$, one can get information about the $\vec{q}_\perp$-dependence of the GTMD. The problem, however, is that the scattering amplitude 
$\vec{M}$ is given by a complicated convolution integral of the dipole $S$-matrix. For the transversely polarized virtual photon, the relation is
\beq
\frac{d\sigma}{d\vec{P}_\perp d\vec{\Delta}_\perp} \propto |\vec{M}|^2\,, \qquad 
\vec{M}(\vec{P}_\perp,\vec{\Delta}_\perp) = \int \frac{d^2\vec{q}_\perp}{2\pi}  \frac{\vec{P}_\perp - \vec{q}_\perp}
{(\vec{P}_\perp-\vec{q}_\perp)^2+\epsilon_f^2} S_Y(\vec{q}_\perp, \vec{\Delta}_\perp) \,, 
\label{q}
\eeq
where $\epsilon^2_f=z(1-z)Q^2+m_f^2$. (Here, $z$ (or $1-z$) is the momentum fraction of the quark (or antiquark) and $Q^2$ is the photon virtuality. 
We neglect the quark mass $m_f$.) In order to make the extraction of $S$ from $M$ easier, the authors of Ref.~\cite{Hatta:2016dxp} 
suggested looking at the small-$Q^2$ region where the $\vec{q}_\perp$-integral in Eq.~(\ref{q}) is dominated by $\vec{q}_\perp\sim \vec{P}_\perp$. 
In this paper, we push this idea to the extreme and consider the photoproduction limit of small $Q^2\to 0$.

In the lepton-nucleon scattering, approaching the kinematical boundary $Q^2\to 0$ is experimentally feasible as HERA indeed has measured the parton density functions (PDFs) 
in the proton down to $Q^2=0.05$ GeV$^2$. There is, however, a more efficient way to prepare a flux of almost real photons. This is $pA$ UPCs in which the nucleus is treated 
only as a source of WW photons. By using a large nucleus, the smallness of the electromagnetic coupling $\alpha_{em}$ is 
compensated by the atomic number squared $Z^2$. Moreover, since the photons are almost on shell, they only have transverse polarizations. (When $Q^2\neq 0$, 
the contribution from the longitudinally polarized virtual photon should be added to Eq~(\ref{q}); see also Refs.~\cite{Chyla:2000ue,Chyla:2000cu}.) 
Note in our case one should ensure exclusivity of the process such that the proton and nucleus remain intact. This is especially important for the proton as one should 
detect the final proton in order to recover the full kinematics necessary for GTMD. 
On the untagged nucleus side  we do not consider resolved photon processes but we rather concentrate on the so-called direct photon process. 
Below we consider exclusive diffractive dijet production in UPCs and demonstrate that, in the ideal case $Q^2=0$, the convolution (\ref{q}) can be analytically inverted.

Let us consider the kinematics of this process first. We choose a frame in which the nucleus moves in the $+z$ direction and the proton moves in the $-z$ direction. 
We are interested in measuring the GTMD distribution of the proton. For this purpose, we require a large rapidity gap in the final state between the recoiling 
proton and the dijet at forward rapidities $y_{1,2}\gg 1$. The invariant mass of the dijet providing the hard scale for the process is then given by
\beq
m^2 = 2(k_1^+ +k_2^+)(k_1^- + k_2^-)-(\vec{k}_{1\perp}+\vec{k}_{2\perp})^2 = 
2k_{1\perp}k_{2\perp}(\cosh(y_1-y_2)-\cos \phi_{12})\,, 
\eeq
where the conventional definition for the light-cone momenta $l^\pm=(l^0\pm l^3)/\sqrt{2}$, $\vec{l}_\perp=(l^1,l^2)$ is adopted.
The relevant value of the proton momentum fraction $x$ is  determined from the condition $(xp+q)^2=m^2$ to be 
\beq
 x=\frac{m^2}{4E_p \omega} = 
\frac{k_{1\perp}k_{2\perp}(\cosh(y_1-y_2)-\cos \phi_{12})}{2E_p \omega}\,,
\eeq
where $q^\mu=\omega(1,0,0,1)$ is the WW photon momentum. The rapidity gap is then 
\beq
\Delta y = {\rm min}\{y_1,y_2\} + \ln \frac{\sqrt{2}p^-}{\Delta_\perp} = {\rm min} \left\{\ln \frac{\sqrt{2}k_1^+}{k_{1\perp}}\,, \;
\ln \frac{\sqrt{2}k_2^+}{k_{2\perp}}   \right\} +\ln \frac{2E_p}{\Delta_\perp} \sim \ln \frac{E_p \omega}{k_\perp^2} \sim \ln \frac{1}{x}=Y\,.
\label{gap}
\eeq

In order to relate $pA$ UPCs to $p\gamma$ collisions, we use the standard formula 
\beq
\frac{d\sigma^{pA}}{dy_1dy_2 d^2\vec{k}_{1\perp}d^2\vec{k}_{2\perp}} = 
\int d\omega \frac{dN}{d\omega} \frac{d\sigma^{p\gamma}}{dy_1dy_2 d^2\vec{k}_{1\perp}d^2\vec{k}_{2\perp}}\,, 
\label{omega}
\eeq 
where the photon flux is given by
\beq
\frac{dN}{d\omega} = \frac{2Z^2\alpha_{em}}{\pi \omega} \left[ \xi K_0(\xi)K_1(\xi)-\frac{\xi^2}{2}(K_1^2(\xi)-K_0^2(\xi))\right]\,,
\eeq
with $\xi = \omega \frac{R_p+R_A}{\gamma}$ and $\gamma = \frac{\sqrt{s_{NN}}}{2m_p}$. Here, $R_p$ and $R_A$ are the radii of the proton 
and the nucleus, respectively. The flux $dN/d\omega$ decays exponentially when $\xi \gg 1$. In the $p+$Au collisions at the RHIC at $\sqrt{s_{NN}}=200$ GeV, 
we have $Z=79$, $\gamma=100$ and $R_A\approx 8$ fm. The characteristic photon energy corresponding to $\xi\sim 1$ is $\omega \sim 2$\,GeV. 
In the $p$+Pb collisions at $\sqrt{s_{NN}}=5.02$ TeV at the LHC, we have instead $Z=82$, $\gamma\approx 2500$ and $\omega \sim 50$ GeV. 
In the UPCs the  virtualities of the emitted photons are small compared to the typical scales of a hard process so they can be considered real
to a good approximation. The $p\gamma$ cross section in Eq.~(\ref{omega}) is calculated as follows \cite{Altinoluk:2015dpi}
\beq
\frac{d\sigma^{p\gamma }}{dy_1dy_2 d^2\vec{k}_{1\perp}d^2\vec{k}_{2\perp}} = 
N_c \alpha_{em}(2\pi)^2 q^+\delta(k^+_1+k^+_2-q^+) \sum_f e_f^2 2z(1-z) (z^2+(1-z)^2) |\vec{M}|^2\,, \label{delta}
\eeq
where $q^+=\sqrt{2}\omega$ and 
\beq
z=\frac{k_{1\perp}e^{y_1}}{ k_{1\perp}e^{y_1} + k_{2\perp}e^{y_2}}\,.
\eeq
The amplitude $\vec{M}$ is given by Eq.~(\ref{q}) with $Q^2\approx 0$
\beq
\vec{M}(\vec{P}_\perp,\vec{\Delta}_\perp) = \int \frac{d^2\vec{q}_\perp}{2\pi}  \frac{\vec{P}_\perp - 
\vec{q}_\perp}{(\vec{P}_\perp-\vec{q}_\perp)^2} S(\vec{q}_\perp, \vec{\Delta}_\perp) \,. \label{qq}
\eeq
(In the following we suppress the subscript $Y$.) 

As discussed in Refs.~\cite{Hatta:2016dxp,Hagiwara:2016kam}, the dominant angular dependence of $S$ is elliptic, namely, we can approximately 
parametrize it as 
\beq
S(\vec{q}_\perp, \vec{\Delta}_\perp) = S_0(q_\perp, \Delta_\perp) +
2\cos 2(\phi_q-\phi_\Delta) \tilde{S}(q_\perp,\Delta_\perp)\,.
\eeq
The angular integral in Eq.~(\ref{qq}) can then be calculated analytically  
\beq
\int_0^\infty q_\perp dq_\perp \int_0^{2\pi} d\phi_q \frac{\vec{P}_\perp - 
\vec{q}_\perp}{(\vec{P}_\perp-\vec{q}_\perp)^2} S_0(q_\perp,\Delta_\perp) = 
\frac{2\pi \vec{P}_\perp}{P_\perp^2}\int_0^{P_\perp} dq_\perp  q_\perp S_0(q_\perp,\Delta_\perp) \,.
\eeq
\beq
\int_0^{2\pi} d\phi_q \frac{\vec{P}_\perp - \vec{q}_\perp}{(\vec{P}_\perp^2-\vec{q}_\perp)^2}
\cos 2(\phi_q-\phi_\Delta) &=& \frac{\pi\vec{P}_\perp}{P_\perp^2} \cos 2(\phi_P-\phi_\Delta) \nonumber \\
&\times& \left( \frac{q_\perp^2}{P_\perp^2}\theta(P_\perp-q_\perp) -\frac{P_\perp^2}{q_\perp^2}
\theta(q_\perp-P_\perp)\right)+\cdots \,, \label{2nd}
\eeq
where in Eq.~(\ref{2nd}) we have omitted the terms orthogonal to $\vec{P}_\perp$ which drop out in the formulas below.
We then trivially perform the $\omega$-integral in Eq.~(\ref{omega}) 
using the $\delta$-function and arrive at 
\beq
\frac{d\sigma^{pA}}{dy_1dy_2 d^2\vec{k}_{1\perp}d^2\vec{k}_{2\perp}} \label{aaaa}   \approx \omega \frac{dN}{d\omega} 
\frac{2(2\pi)^4N_c \alpha_{em}}{P_\perp^2} 
\sum_f e_f^2 z(1-z) (z^2+(1-z)^2) \bigl(A^2+2\cos 2(\phi_P-\phi_\Delta) AB \bigr)\,,  \nonumber \\
\label{fin}
\eeq
where we defined 
\beq
A(P_\perp,\Delta_\perp)\equiv - \int_0^{P_\perp} dq_\perp  q_\perp S_0(q_\perp,\Delta_\perp)\,, 
\label{ad}
\eeq
\beq
B(P_\perp,\Delta_\perp) \equiv -\int_0^{P_\perp} dq_\perp \frac{q_\perp^3}{P_\perp^2}\tilde{S}(q_\perp,\Delta_\perp) +
\int_{P_\perp}^\infty  dq_\perp \frac{P_\perp^2}{q_\perp} \tilde{S}(q_\perp,\Delta_\perp)\,.
\label{bd}
\eeq
In (\ref{fin}), it is understood that
\beq
\omega=\frac{1}{\sqrt{2}}(k_1^+ + k_2^+) = \frac{1}{2}(k_{\perp 1}e^{y_1} + k_{\perp 2} e^{y_2}) \,,
\eeq
and we have neglected the quadratic terms $\tilde{S}^2$ because the magnitude of $\tilde{S}$ is at most a few percent of 
that of $S_0$ \cite{Hagiwara:2016kam}.

By a Fourier analysis, experimentalists can extract $A$ and $B$ from the dijet data as  functions of $P_\perp$ and $\Delta_\perp$. 
(Note that the overall sign of $A$ and $B$ cannot be determined. But this can be fixed by comparing with model predictions, see below.) 
From this, one can easily reconstruct $S_0(P_\perp,\Delta_\perp)$ via
\beq
S_0(P_\perp,\Delta_\perp) =-\frac{1}{P_\perp} \frac{\partial}{\partial P_\perp}A(P_\perp,\Delta_\perp)\,.
\eeq
This is a very direct determination of the dipole $S$-matrix. Reconstructing the elliptic part $\tilde{S}$ by inverting 
the relation (\ref{bd}) is more involved.  
Let us write $P_\perp^2=e^v$ and $q_\perp^2=e^u$. Then Eq.~(\ref{bd}) takes the following form 
(suppressing $\Delta_\perp$ for the moment)
\beq
B(v) = \int_{-\infty}^\infty du \left(-e^{u-v}\theta(v-u)+ e^{v-u}\theta(u-v)\right) C(u)\,, \label{ba}
\eeq
where we defined $C(u)=\frac{q_\perp^2}{2}\tilde{S}(q_\perp)$. The expression in Eq.~(\ref{ba}) can be deconvoluted by Fourier-transforming in $v$, 
but there is a more direct method. By a further change of variables, one can write Eq.~(\ref{ba}) as  
\beq
B(v) = \int_{0}^\infty dt\,  e^{-t} \left(C(v+t)- C(v-t)\right)\,. \label{badif}
\eeq
Expanding the difference in brackets into Taylor series around the symmetric point, one gets
\beq
B(v) =  2 \sum_{k=0}^{\infty} \frac {d^{2k+1}}{dv^{2k+1}} C(v)\,. \label{baser}
\eeq
From this one can easily obtain 
\beq
\frac{d C(v)}{dv} = \frac{1}{2} \left(B(v)- \frac{d^2 B(v)}{dv^2}\right)\,, \label{barec}
\eeq
 or in the original variables,
\beq
\tilde{S}(P_\perp,\Delta_\perp) = -\frac{\partial B(P_\perp,\Delta_\perp)}{\partial P_\perp^2} +
\frac{2}{P_\perp^2}\int^{P_\perp^2}_0 \frac{dP'^2_\perp}{P'^2_\perp} B(P'_\perp, \Delta_\perp)\,.
\eeq
We thus find that both $S_0$ and $\tilde{S}$ can be fully reconstructed from the experimental data. Via Eq.~(\ref{st}), this can be used 
to determine the gluon GTMD distribution. Its $x$-dependence can also be studied by measuring the cross section as a function of 
the rapidity gap $\Delta y \approx Y$, see Eq.~(\ref{gap}). Later we discuss the experimental feasibility of such a measurement. 
\begin{figure}[!h]
 \includegraphics[width=75mm]{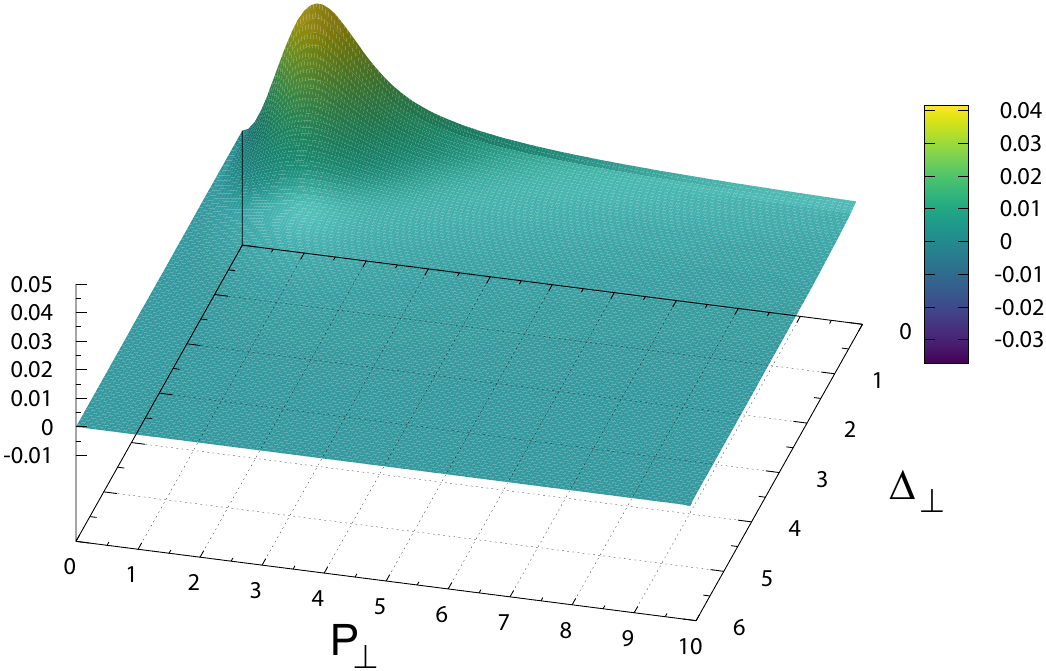}
 \includegraphics[width=75mm]{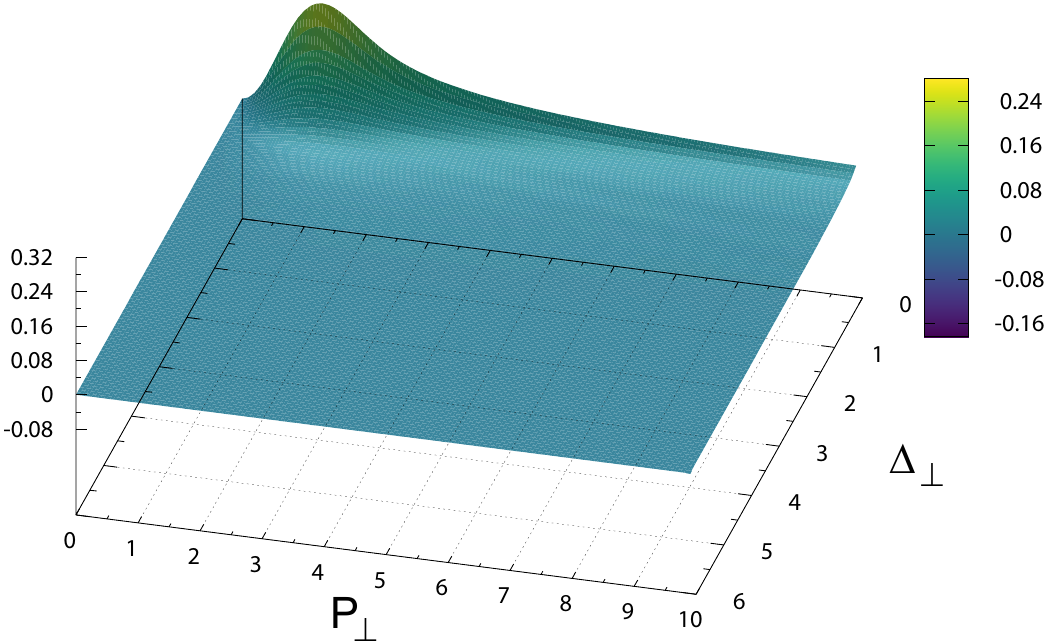}
\caption{ $A(P_\perp,\Delta_\perp)$ at $Y=4$ (left) and $Y=8$ (right). Here, both $P_\perp$ and $\Delta_\perp$ are given 
	in units of $1/R$, $R=0.4$ fm.} 
\label{bk1}
\end{figure}
\begin{figure}[!h]
	\includegraphics[width=75mm]{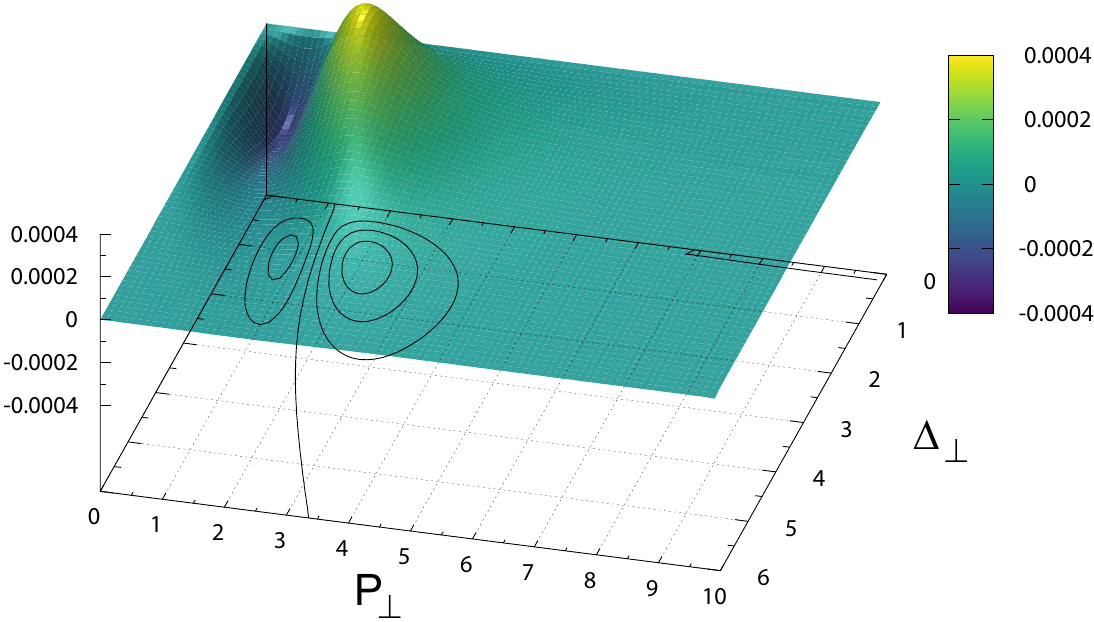}
	\includegraphics[width=75mm]{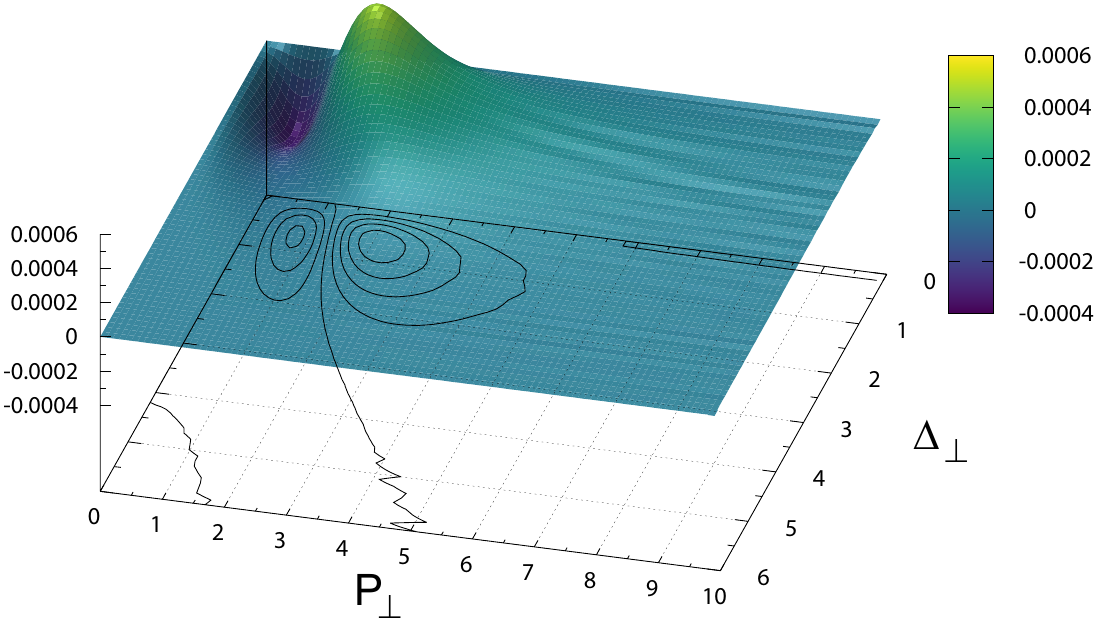}
	\caption{$B(P_\perp,\Delta_\perp)$ at $Y=4$ (left) and $Y=8$ (right). Here, both $P_\perp$ and $\Delta_\perp$ are given 
	in units of $1/R$, $R=0.4$ fm.} 
\label{bk2}
\end{figure}

\section{Model calculations}
\label{Sec:model-calc}

In this section, we provide theoretical inputs and model calculations for $A$ and $B$ which  hopefully can serve as a guidance for 
the experimental measurement of these functions. On general grounds, we expect the following asymptotics
\beq
S_0(q_\perp,\Delta_\perp) \sim \begin{cases} {\rm const} & q_\perp\to 0 \\ \left(\frac{1}{q_\perp}\right)^4 & q_\perp \to \infty\,,
\end{cases} \qquad \tilde{S}(q_\perp,\Delta_\perp) \sim \begin{cases} q_\perp^2 & q_\perp\to 0 \\ \left(\frac{1}{q_\perp}\right)^6 & q_\perp \to \infty\,.
\end{cases}
\eeq
The behavior of $S_0$ at large $q_\perp$ follows from dimensional analysis while that of $\tilde{S}$ requires some explanations. 
At small-$q_\perp$, the factor $q_\perp^2$ comes from the $J_2$ Bessel function associated with the $\cos 2\phi$ distribution. 
In coordinate space, the elliptic part $\tilde{S}(r_\perp,b_\perp)$ has an extra factor $r_\perp^2$ compared to $S_0(r_\perp,b_\perp)$, 
see the discussion around Eq.~(22) of Ref.~\cite{Hagiwara:2016kam}. In momentum space, this gives an extra factor of $1/q_\perp^2$ 
at large-$q_\perp$. We thus deduce that
\beq
B(P_\perp) = -\frac{1}{P_\perp^2} \int_0^{P_\perp} dq_\perp q_\perp^3 \tilde{S}(q_\perp) +
P_\perp^2\int_{P_\perp}^\infty dq_\perp \frac{\tilde{S}(q_\perp)}{q_\perp} \sim 
\begin{cases} P_\perp^2 & P_\perp\to 0 \\  \left(\frac{1}{P_\perp}\right)^2 & P_\perp \to \infty\,.
\end{cases}
\eeq
\begin{figure}[!h]
	\includegraphics[width=75mm]{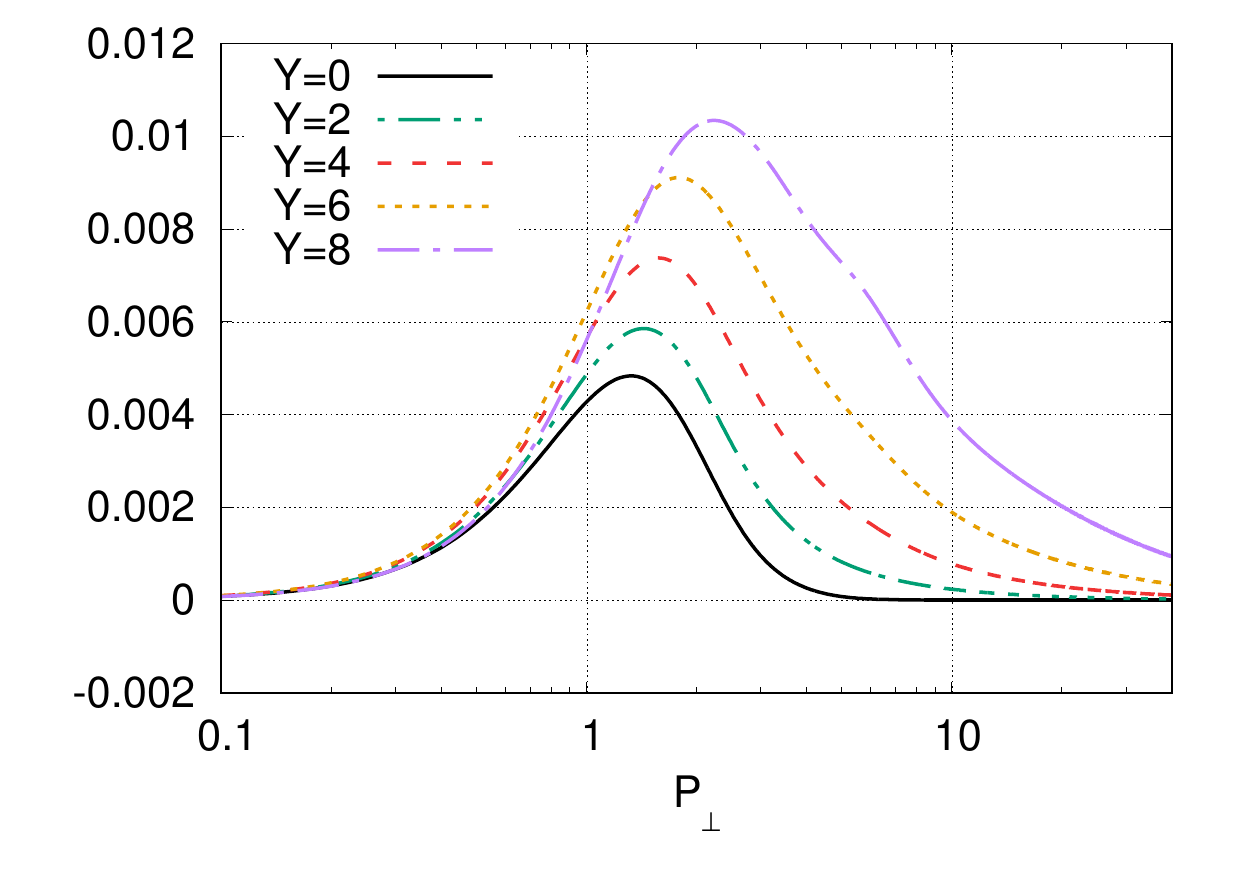}
	\includegraphics[width=75mm]{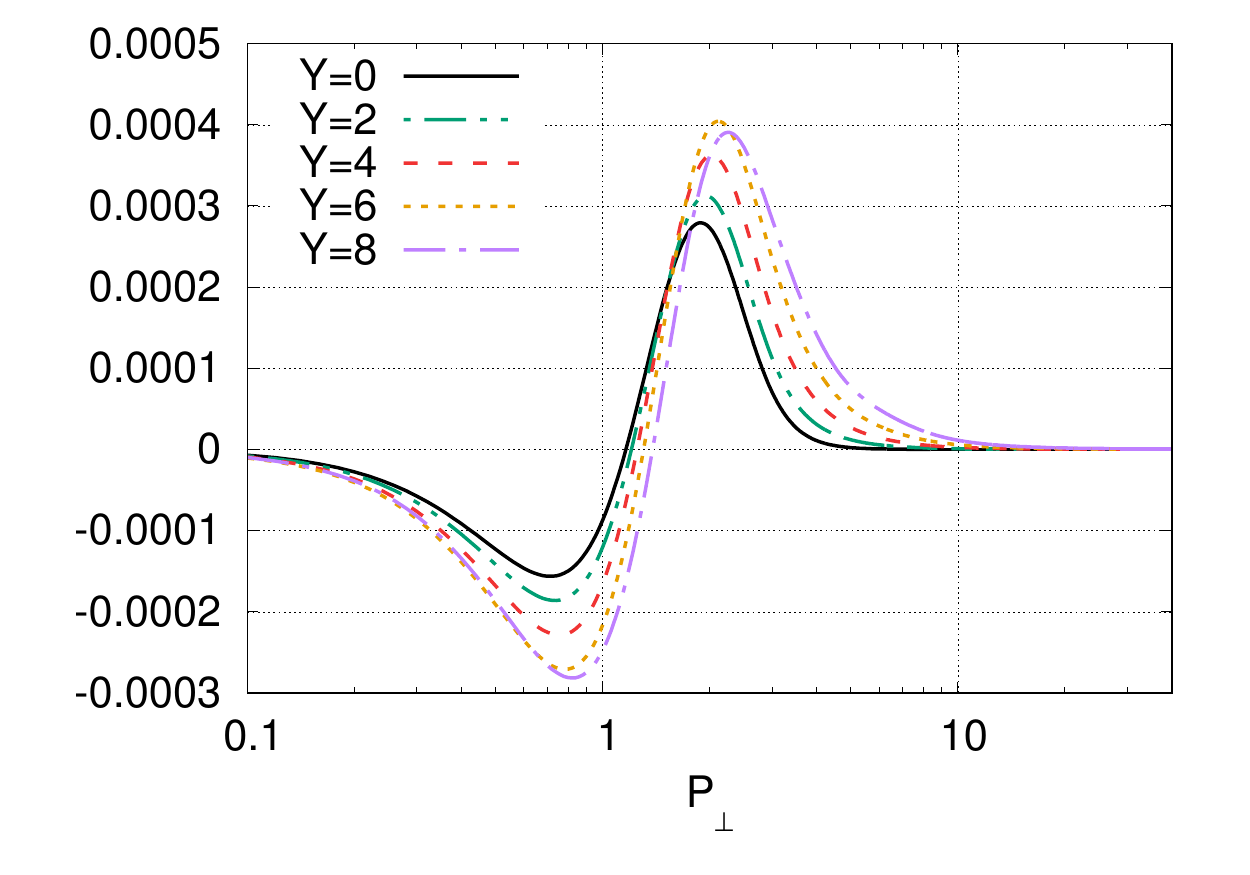}
	\caption{The rapidity evolution of $A$ (left) and $B$ (right) at $\Delta_\perp = 1.0$. Here, both $P_\perp$ and $\Delta_\perp$ are given 
	in units of $1/R$, $R=0.4$ fm.} 
\label{bk3}
\end{figure}

We now compute $A$ and $B$ for both  proton and nuclear targets using the BK model and the MV model, respectively. In the first case, we use a numerical 
solution of the BK equation with impact parameter obtained in \cite{Hagiwara:2016kam}. 
For technical reasons, in what follows we switch from the $S$-matrix to the $T$-matrix as $S=1-T\to -T$. (The `1' term does not contribute when 
$\vec{\Delta}_\perp\neq 0$, but in numerical integrations with a finite momentum cutoff this can cause unphysical oscillations.) In Figs.~\ref{bk1} and \ref{bk2}, 
we show the numerical results as 3D plots of $A(P_\perp,\Delta_\perp)$ and $B(P_\perp,\Delta_\perp)$ functions, respectively, at 
$Y=\ln \frac{1}{x}=4$ (left) and $Y=8$ (right). The model contains a single parameter (denoted $R$ in Ref.~\cite{Hagiwara:2016kam}) which sets 
the typical length scale. For a realistic initial condition we choose $R=0.4$ fm, which means that $P_\perp=1$ in these plots should be interpreted 
as $P_\perp=0.5$ GeV. In Fig.~\ref{bk3} we show the $Y$-dependence of $A(P_\perp)$ and $B(P_\perp)$ at fixed $\Delta_\perp=1$. As one can expect 
from the findings in Ref.~\cite{Hagiwara:2016kam}, the location of the peak in $A$, which is identified with the saturation momentum $Q_s(Y)$ moves toward 
the larger $P_\perp$ region at a constant ``speed'' representing the geometric scaling, whereas that of the elliptic part moves very slowly with $Y$. The rapidity 
dependence of the maximum peak momentum of $A$ at fixed $\Delta_\perp$ seems the same as the peak momentum of the Wigner distribution at a fixed 
impact parameter depicted in Ref.~\cite{Hagiwara:2016kam}.
\begin{figure}[!h]
	\includegraphics[width=75mm]{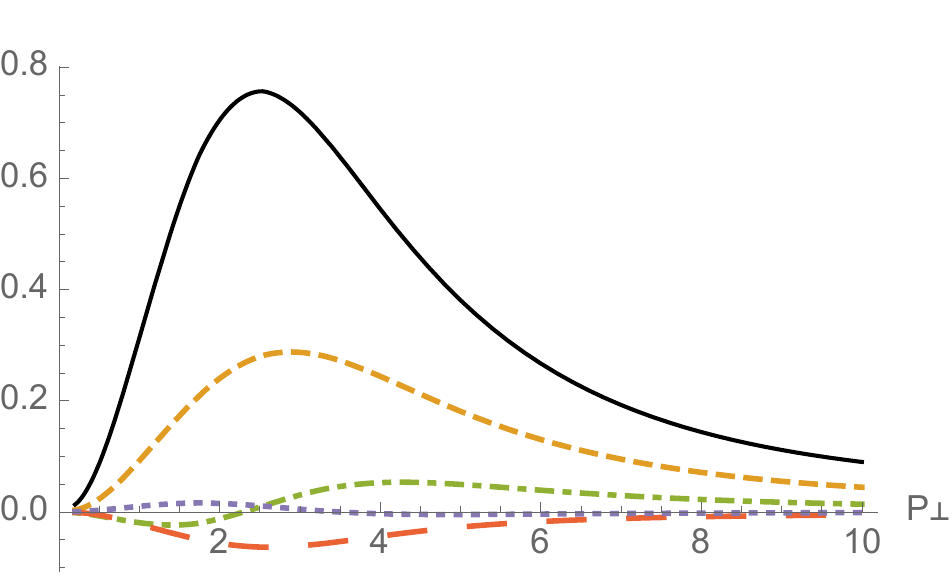}
	\includegraphics[width=75mm]{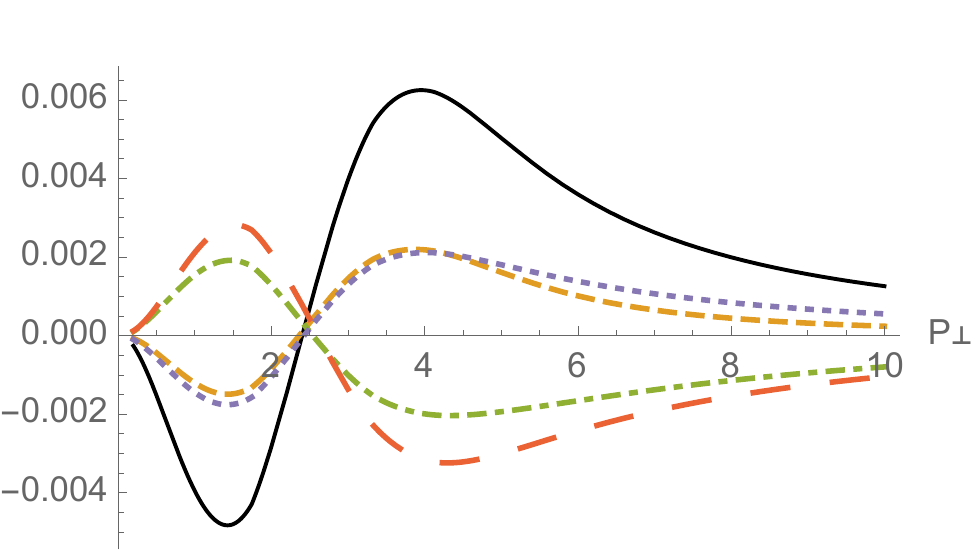}
	\caption{The $P_\perp$ dependence of $A$ (left) and $B$ (right) in the MV model for a nuclear target ($A=208$) following Ref.~\cite{Iancu:2017fzn} 
	at $\Delta_\perp = 0.2$ (solid lines), $\Delta_\perp = 0.25$ (dashed lines), $\Delta_\perp = 0.3$ (dotted-dashed lines), $\Delta_\perp = 0.4$ (long dashed lines), 
	and $\Delta_\perp = 0.5$ (dotted lines). Here, both $P_\perp$ and $\Delta_\perp$ are given 
	in units of $1/R$, $R=0.4$ fm.
} \label{mv1}
\end{figure}
\begin{figure}[!h]
	\includegraphics[width=75mm]{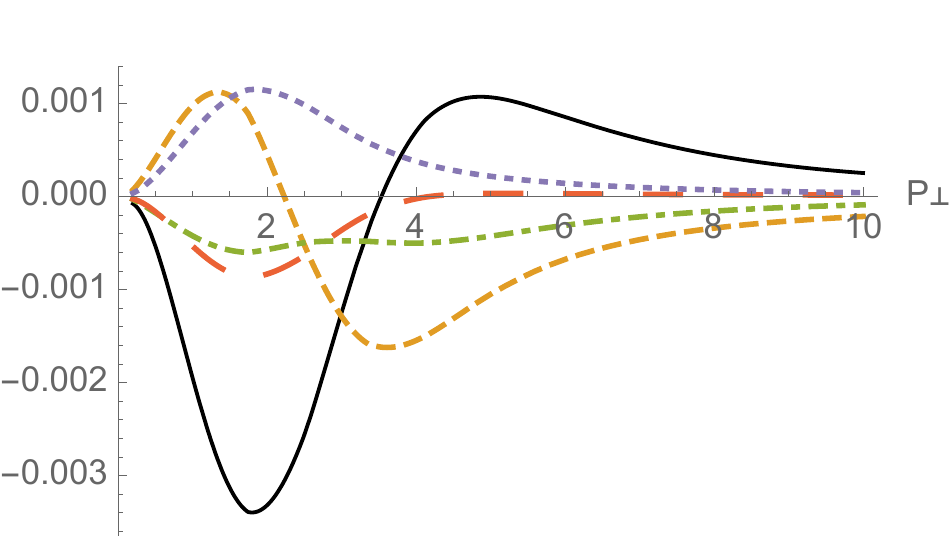}
	\includegraphics[width=75mm]{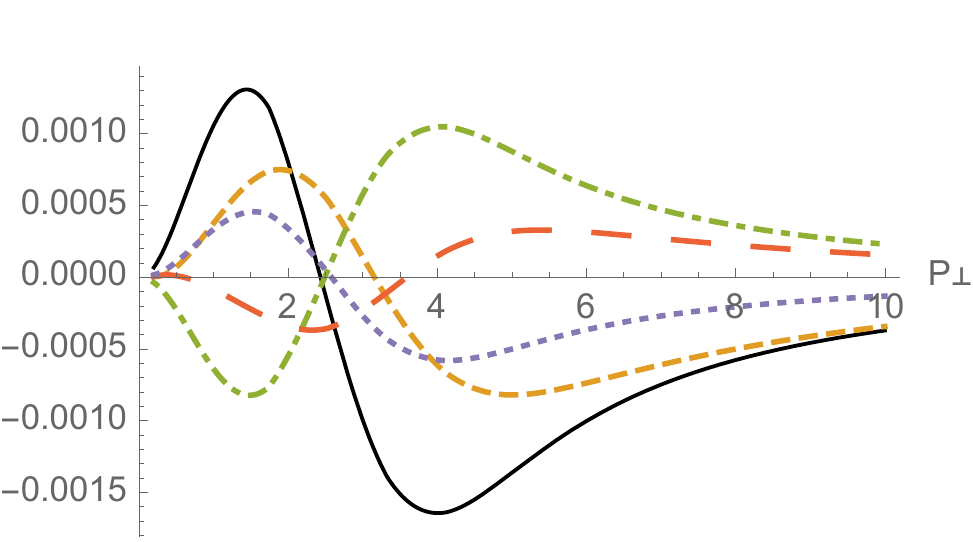}
	\caption{The $P_\perp$ dependence of $A$ (left) and $B$ (right) in the MV model for a nuclear target ($A=208$) following Ref.~\cite{Iancu:2017fzn} 
	at $\Delta_\perp = 0.7$ (solid lines), $\Delta_\perp = 0.8$ (dashed lines), $\Delta_\perp = 0.9$ (dotted-dashed lines), $\Delta_\perp = 1.0$ (long dashed lines), 
	and $\Delta_\perp = 1.1$ (dotted lines). Here, both $P_\perp$ and $\Delta_\perp$ are given 
	in units of $1/R$, $R=0.4$ fm.
} \label{mv2}
\end{figure}

Next we compute $A$ and $B$ for a  nuclear target. (Though our primary interest is UPCs in $pA$ collisions, one can also consider measuring the gluon Wigner 
distribution of a nucleus from UPCs in $AA$ collisions.) In the MV model, the analytical expressions of $S_0(r_\perp,b_\perp)$ and $\tilde{S}(r_\perp,b_\perp)$ 
for the nuclear target in the impact parameter space have recently become available in terms of the nuclear thickness function $T_A(b)$ and its derivatives 
\cite{Iancu:2017fzn} (for earlier calculations, see Ref.~\cite{Zhou:2016rnt}). We have not implemented $x$-dependence in this model. The nuclear saturation 
scale used in our calculations scales as $Q_{s,A}^2=A^{1/3}Q_{0s}^2$ in terms of the proton saturation momentum at zeroth impact parameter taken to 
be $Q_{0s}^2=R^{-2}$, $R=0.4$ fm, and the nuclear radius is $R_A=(1.12\,{\rm fm})A^{1/3}$. For a large nucleus with $A=208$, we have performed 
the Fourier transforms $\{\vec{r}_\perp,\vec{b}_\perp\}\to \{\vec{q}_\perp,\vec{\Delta}_\perp\}$ and numerically evaluated the integrals in 
Eqs.~(\ref{ad}) and (\ref{bd}). The Gaussian weights ${\rm exp}(-\epsilon_r r^2)$ and ${\rm exp}(-\epsilon_b b^2)$ (with $\epsilon_r=(0.5\,{\rm fm})^{-2}$ and 
$\epsilon_b=R_A^{-2}$), effectively cutting off effects at large $\{\vec{r}_\perp,\vec{b}_\perp\}$ in the corresponding Fourier integrals, have been used 
to ensure that no unphysical contributions and oscillations arise for large enough limits of integration. 

The results for $A(P_\perp,\Delta_\perp)$ and $B(P_\perp,\Delta_\perp)$ are shown in Figs.~\ref{mv1} and \ref{mv2} (left and right panels, respectively). While at small 
$\Delta_\perp<0.3$ the function $A(P_\perp,\Delta_\perp)$ is large and positive definite, it quickly vanishes at large $\Delta_\perp$ and $P_\perp$ where 
the function changes its sign and a single node appears. The elliptic contribution $B(P_\perp,\Delta_\perp)$ also exhibits a single node and a slower 
dependence on $\Delta_\perp$. For any $\Delta_\perp$, it has a node in the $P_\perp$ dimension whose position is almost independent of $\Delta_\perp$. 
Similarly to $A$, the function $B$ vanishes at large $P_\perp$. In the case of large nuclear targets, typical peak values of both $A$ and $B$ at low 
$\Delta_\perp<0.3$ are 1-2 orders of magnitude larger than those for the proton target found in the BK model. At larger $\Delta_\perp$, 
the function $A$ disappears very quickly and becomes smaller than that of the proton. Interestingly enough, at $\Delta_\perp>0.7$ the MV model
in the nuclear case predicts an order of magnitude larger $B$ than that in the proton case. This means that the elliptic Wigner distribution can be
better constrained by measuring the exclusive dijet production cross section in $AA$ UPCs. Clearly, both the BK and MV model predictions exhibit 
nontrivial dynamics in the low-$P_\perp<4$ GeV region while they are smoothly and monotonously vanishing at larger $P_\perp$, making it difficult to probe 
the corresponding features at the LHC.

\section{Prospects for Wigner distribution measurements}
\label{Sec:Exp}

In what follows, we study the key ingredients of the Wigner function by measuring the functions $A$ and $B$. The process of interest is exclusive 
diffractive dijet production in $pA$ UPCs where the dijet system goes in the direction of the nucleus, while on the opposite side, the intact proton is measured. 

The event selection is based on requiring a dijet system which should ideally be accompanied by large gaps at both, the photon and proton sides (see e.g. the ATLAS 
analysis of multijet events in PbPb data \cite{ATLAS_dijetpPb}). The photon can, in principle, develop its structure but we are interested in the exclusive initial state, 
i.e. the direct photon process; thus, we have to require an isolated photon and hereby suppress events with photon remnants.
One has to require a precise measurement of the intact proton and possibly of the intact nucleus, 
see the discussion below. As is seen from Eq.~(\ref{fin}), if we integrate over all $\phi_P$ and $\phi_\Delta$ angles, the second term proportional to $AB$ disappears, 
and then we only probe $A$ as a function of $P_\perp$ and $\Delta_\perp$, and hence $S_0$, according to Eq.~(\ref{ad}).

Therefore it would be of primary importance to measure the $\phi$-integrated cross section separately as a function of $P_\perp$, while keeping $\Delta_\perp$
integrated out within its typical detector acceptance window, and vice versa, as a function of $\Delta_\perp$ integrating it over $P_\perp$. 

First, as we know from diffraction measurements at HERA and Tevatron, the cross section falls roughly exponentially with $|t| = \Delta_\perp^2$ 
as $\exp(-b|t|)$ with a slope $b \approx$ 7 GeV$^{-2}$. This is one of the key observables used in the context of constraining the dipole cross section or unintegrated 
gluon PDFs from the data. At LHC, the $t$ variable is measured using special forward proton detectors TOTEM (at the CMS side) or ALFA (at the ATLAS side) with 
relatively small systematic uncertainties. It can also be measured using e.g.~the sum of the jet transverse momenta found
in the central detector but with a worse resolution and a larger systematic uncertainty compared to forward proton detectors. The $t$-dependence was indeed 
measured for the elastic cross section in $pp$ collisions rather thoroughly using TOTEM (see for example Refs.~\cite{Antchev:2013gaa,Antchev:2013paa}) 
and ATLAS (see Refs.~\cite{Aad:2014dca,Aaboud:2016ijx}). The $t$-distribution at the proton vertex for $p$+Pb collisions has not yet been measured but 
the data samples collected in $p$+Pb runs where intact protons are tagged by TOTEM or ALFA detectors exist and are ready to be analyzed.
Very recently other sets of forward proton detectors such as AFP \cite{AFP,Tasevsky:2015xya} (ATLAS forward proton) and CT-PPS \cite{CT-PPS} 
(CMS-TOTEM Precision Proton Spectrometer) have been installed as well. Their sensors are radiation hard enough to be inserted in the LHC beam 
at all times and they also dispose of fast time-of-flight detectors to suppress pileup background. Consequently, the broad physics program 
including measuring diffractive and exclusive processes in $pp$ and heavy ion collisions can be pursued. We are aware that while the physics case for the use of forward proton
spectrometers in the $pp$ collisions has been thoroughly studied during recent years, feasibility studies (such as acceptance, event yield and background rejection) for using these detectors in the heavy ion collisions still need to be done. The central system in $p$+Pb collisions is expected to be shifted by $\Delta y \approx 0.46$ with respect to that 
in $pp$ collisions which seems to still be manageable. If the detection of the intact nucleus turns out to be too difficult, one can require the Zero-Degree Calorimeter to be 
empty in the direction of the photon. With results presented in this text, we hope to encourage and motivate experimentalists to perform such studies.

Second, the $P_\perp$ variable carries the information about the hard scale of the event since it is identified with the transverse momentum of the dijet system.
At LHC jets are usually measured above $P_\perp$ of 15--20~GeV because below these values, the resolution exceeds 10\% and measurements suffer from large
systematic uncertainties. The results shown in Fig.~\ref{bk3}, however, indicate that the most visible saturation effects in the Wigner function are observed for 
$P_{\perp} \lesssim$ 10~GeV. In this respect, $pA$ data from the RHIC could be useful since jets with $P_{\perp} > 5$~GeV can be measured rather reliably there. 
The Roman pots in the PP2PP detector \cite{PP2PP} measured intact protons only in $p+$Al and $p+$Au runs, while ions could not be measured because of large backgrounds.

If we are able to measure $A(P_\perp, \Delta_\perp)$ in the way described above, the $B(P_\perp, \Delta_\perp)$ and hence $\tilde{S}$ should be reachable by measuring
$(\phi_P-\phi_\Delta)$-dependence of the cross section in bins of $P_\perp$ and $\Delta_\perp$ since its amplitude is directly related to the product $AB$.

To access the $x$-dependence of the Wigner function (or that of $S_0$ and $\tilde{S}$ functions) one has to measure both the $\phi$-integrated and
$\phi$-dependent cross sections as functions of jet rapidity difference $(y_1-y_2)$ in fixed bins of $P_\perp$ and $\Delta_\perp$. Such a mapping could give us 
almost complete information about the Wigner distribution within the accessible kinematic windows. 

\section{Conclusions}
\label{Sec:Conclusions}

To summarize, our work suggests and explores a new potentially important way to constrain the gluon Wigner distribution by measuring the exclusive diffractive dijet 
production cross section in $pA$ (and also in $AA$) UPCs at high energies. We demonstrate that both components of the gluon Wigner distribution may be extracted 
from these data. The special role here is played by the elliptic component of the Wigner (or the corresponding GTMD) function providing 
the $\cos 2\phi$ dependence on the angle $\phi$ between the jets total and relative transverse momenta. 

The angular dependence of the Wigner function was recently shown \cite{Hagiwara:2017ofm,Iancu:2017fzn} to be a complementary way to describe an elliptic flow 
in $pA$ collisions, the gluon transversity GPD and angular correlation in DVCS \cite{Hatta:2017cte} as well as the angular correlation in quasi-elastic scattering 
$\gamma_T^*A\to A'X$ on a nucleus $A$ \cite{Zhou:2016rnt}. It is of interest how this mechanism should be combined with the ``standard'' collective mechanism 
of elliptic flow generation in quark-gluon plasma. In the pioneering studies of elliptic flow of direct photons \cite{Kopeliovich:2007fv,Kopeliovich:2007sd} and pions 
\cite{Kopeliovich:2008nx} in the dipole approach this mechanism was considered as an additive to the standard one. At the same time, as explained in a recent 
paper \cite{Gotsman:2016whc}, azimuthal long-range rapidity correlations in proton-proton collisions stem from the CGC/saturation physics, and not from quark-gluon 
plasma production, implying rather a sort of complementarity between dipole and plasma mechanisms. There are also other attempts to describe the flowlike effects 
as providing a complementary picture to the hydrodynamical description, for example using the rope hadronization \cite{Bierlich:2016vgw}. Notably, such a relation to 
the elliptic flow may also be the manifestation of complementarity (or ``duality'') between statistical and dynamical descriptions of heavy-ion collisions \cite{Cleymans:2011zz}.

Our study also aims to encourage experimentalists to look in more detail to see if the Wigner function could be measured at the RHIC, LHC and EIC. We tried to provide first experimental 
ideas but more work would be needed before making conclusions. It is encouraging that data which could be used for such studies exist. There are samples of $pA$ data 
at both, the LHC and RHIC, where forward protons were tagged by Roman pots. The RHIC environment seems to be more suitable to look at saturation effects that are expected 
to be visible at $p_T<$ 10~GeV, while the LHC data promise to provide more accurate measurements of protons and jets.

Finally, we remind the reader that the Wigner distribution can also be studied in lepton-nucleon scattering at the EIC, as originally suggested in Ref.~\cite{Hatta:2016dxp}. 
By taking the photoproduction limit $Q^2\to 0$, the same functions $A$ and $B$ should come into play. While the process does not enjoy the $Z^2$-enhancement, 
this may be compensated for by the high luminosity achieved at the EIC. 

\section*{Acknowledgments}

Y.~Hatta and O.~Teryaev thank KEK, where this collaboration was initiated, for hospitality and travel support. O.~Teryaev is grateful to JSPS for 
the support of his visit to KEK. He also thanks Institute of Physics, Academy of Sciences of the Czech Republic, for hospitality. Y.~Hagiwara is 
supported by the JSPS KAKENHI Grant No. 17J08072. R.~Pasechnik is partially supported by the Swedish Research Council, 
Contract No.~621-2013-428 and by the Comisi\'{o}n Nacional de Investigaci\'{o}n Cient\'{\i}fica y Tecnol\'{o}gica (CONICYT) 
Project No.~PIA ACT1406. M.~Tasevsky is supported by the Projects No. LG15052 and No. LM2015058 of the Ministry of Education of 
Czech Republic. The numerical calculations were carried out on XC40 at YITP in Kyoto University.


\begin{thebibliography}{99}

\bibitem{Ji:2003ak} 
  X.~d.~Ji,
  Phys.\ Rev.\ Lett.\  {\bf 91}, 062001 (2003).

\bibitem{Belitsky:2003nz} 
  A.~V.~Belitsky, X.~d.~Ji and F.~Yuan,
  Phys.\ Rev.\ D {\bf 69}, 074014 (2004).

\bibitem{Lorce:2011kd} 
  C.~Lorce and B.~Pasquini,
  Phys.\ Rev.\ D {\bf 84}, 014015 (2011).

\bibitem{Meissner:2009ww} 
  S.~Meissner, A.~Metz and M.~Schlegel,
  JHEP {\bf 08}, 056 (2009).

\bibitem{Hatta:2011ku} 
  Y.~Hatta,
  Phys.\ Lett.\ B {\bf 708}, 186 (2012).

\bibitem{Lorce:2013pza} 
  C.~Lorc\'e and B.~Pasquini,
  JHEP {\bf 09}, 138 (2013).

\bibitem{Echevarria:2016mrc} 
  M.~G.~Echevarria, A.~Idilbi, K.~Kanazawa, C.~Lorc\'e, A.~Metz, B.~Pasquini and M.~Schlegel,
  Phys.\ Lett.\ B {\bf 759}, 336 (2016).

\bibitem{Boer:2011fh} 
  D.~Boer {\it et al.},
  arXiv:1108.1713 [nucl-th].

\bibitem{Accardi:2012qut} 
  A.~Accardi {\it et al.},
  Eur.\ Phys.\ J.\ A {\bf 52}, no. 9, 268 (2016).

\bibitem{Hatta:2016dxp} 
  Y.~Hatta, B.~W.~Xiao and F.~Yuan,
  Phys.\ Rev.\ Lett.\  {\bf 116}, no. 20, 202301 (2016).

\bibitem{Altinoluk:2015dpi} 
  T.~Altinoluk, N.~Armesto, G.~Beuf and A.~H.~Rezaeian,
  Phys.\ Lett.\ B {\bf 758}, 373 (2016).

\bibitem{Bhattacharya:2017bvs} 
  S.~Bhattacharya, A.~Metz and J.~Zhou,
  Phys.\ Lett.\ B {\bf 771}, 396 (2017).

\bibitem{Gelis:2010nm} 
  F.~Gelis, E.~Iancu, J.~Jalilian-Marian and R.~Venugopalan,
  Ann.\ Rev.\ Nucl.\ Part.\ Sci.\  {\bf 60}, 463 (2010).

\bibitem{Hagiwara:2016kam} 
  Y.~Hagiwara, Y.~Hatta and T.~Ueda,
  Phys.\ Rev.\ D {\bf 94}, no. 9, 094036 (2016).

\bibitem{Zhou:2016rnt} 
  J.~Zhou,
  Phys.\ Rev.\ D {\bf 94}, no. 11, 114017 (2016).

\bibitem{vonWeizsacker:1934nji} 
  C.~F.~von Weizsacker,
  Z.\ Phys.\  {\bf 88}, 612 (1934).

\bibitem{Williams:1934ad} 
  E.~J.~Williams,
  Phys.\ Rev.\  {\bf 45}, 729 (1934).

\bibitem{Anelli:2008zza} 
  G.~Anelli {\it et al.} [TOTEM Collaboration],
  JINST {\bf 3}, S08007 (2008).

\bibitem{CT-PPS} 
  M. Albrow {\it et al.}, Reports No.~CERN-LHCC-2014-021; No.~TOTEM-TDR-003; No.~CMS-TDR-13.

\bibitem{ALFA} 
  ATLAS Collaboration, Report No.~CERN/LHCC/2008-004.   

\bibitem{AFP} 
  L. Adamczyk {\it et al.}, Reports No.~CERN-LHCC-2015-009; No.~ATLAS-TDR-024;

\bibitem{Tasevsky:2015xya}
  M.~Ta\v{s}evsk\'{y} [ATLAS Collaboration],
  AIP Conf.\ Proc.\  {\bf 1654} 090001 (2015).

\bibitem{Bertulani:2005ru} 
  C.~A.~Bertulani, S.~R.~Klein and J.~Nystrand,
  Ann.\ Rev.\ Nucl.\ Part.\ Sci.\  {\bf 55}, 271 (2005).

\bibitem{Salgado:2011wc} 
  C.~A.~Salgado {\it et al.},
  J.\ Phys.\ G {\bf 39}, 015010 (2012).

\bibitem{Guzey:2016tek} 
  V.~Guzey and M.~Klasen,
  JHEP {\bf 04}, 158 (2016).

\bibitem{Kotko:2017oxg} 
  P.~Kotko, K.~Kutak, S.~Sapeta, A.~M.~Stasto and M.~Strikman,
  Eur.\ Phys.\ J.\ C {\bf 77}, no. 5, 353 (2017).

\bibitem{McLerran:1993ka} 
  L.~D.~McLerran and R.~Venugopalan,
  Phys.\ Rev.\ D {\bf 49}, 3352 (1994).

\bibitem{Iancu:2017fzn} 
  E.~Iancu and A.~H.~Rezaeian,
  Phys.\ Rev.\ D {\bf 95}, no. 9, 094003 (2017).

\bibitem{Balitsky:1995ub} 
  I.~Balitsky,
  Nucl.\ Phys.\ B {\bf 463}, 99 (1996).

\bibitem{Kovchegov:1999yj} 
  Y.~V.~Kovchegov,
  Phys.\ Rev.\ D {\bf 60}, 034008 (1999).

 \bibitem{Chyla:2000ue}
  J.~Chyla and M.~Tasevsky,
  Eur.\ Phys.\ J.\ C {\bf 18}, 723 (2001).
 
  \bibitem{Chyla:2000cu}
  J.~Chyla and M.~Tasevsky,
  Eur.\ Phys.\ J.\ C {\bf 16}, 471 (2000).

\bibitem{ATLAS_dijetpPb} 
ATLAS Collaboration, Report No.~ATLAS-CONF-2017-011.

\bibitem{Antchev:2013gaa}
  G.~Antchev {\it et al.} [TOTEM Collaboration],
  Europhys.\ Lett.\  {\bf 101}, 21002 (2013).

\bibitem{Antchev:2013paa}
  G.~Antchev {\it et al.} [TOTEM Collaboration],
  Phys.\ Rev.\ Lett.\  {\bf 111}, 012001 (2013).

\bibitem{Aad:2014dca}
  G.~Aad {\it et al.} [ATLAS Collaboration],
  Nucl.\ Phys.\ B {\bf 889} 486 (2014).

\bibitem{Aaboud:2016ijx}
  M.~Aaboud {\it et al.} [ATLAS Collaboration],
  Phys.\ Lett.\ B {\bf 761} (2016) 158.

\bibitem{PP2PP} 
 S. Bueltman et al., Nucl.~Instr.~Method in Phys. Research A {\bf 535}, 415 (2004).

\bibitem{Hagiwara:2017ofm} 
  Y.~Hagiwara, Y.~Hatta, B.~W.~Xiao and F.~Yuan,
  Phys.\ Lett.\ B {\bf 771}, 374 (2017).

\bibitem{Hatta:2017cte} 
  Y.~Hatta, B.~W.~Xiao and F.~Yuan,
  Phys.\ Rev.\ D {\bf 95}, no. 11, 114026 (2017).

\bibitem{Kopeliovich:2007fv} 
  B.~Z.~Kopeliovich, H.~J.~Pirner, A.~H.~Rezaeian and I.~Schmidt,
  Phys.\ Rev.\ D {\bf 77}, 034011 (2008).

\bibitem{Kopeliovich:2007sd} 
  B.~Z.~Kopeliovich, A.~H.~Rezaeian and I.~Schmidt,
  Nucl.\ Phys.\ A {\bf 807}, 61 (2008).

\bibitem{Kopeliovich:2008nx} 
  B.~Z.~Kopeliovich, A.~H.~Rezaeian and I.~Schmidt,
  Phys.\ Rev.\ D {\bf 78}, 114009 (2008).

\bibitem{Gotsman:2016whc} 
  E.~Gotsman, E.~Levin, U.~Maor and S.~Tapia,
  Phys.\ Rev.\ D {\bf 93}, no. 7, 074029 (2016).

\bibitem{Bierlich:2016vgw} 
  C.~Bierlich, G.~Gustafson and L.~L\"onnblad,
  arXiv:1612.05132 [hep-ph].

\bibitem{Cleymans:2011zz} 
  J.~Cleymans, G.~I.~Lykasov, A.~S.~Sorin and O.~V.~Teryaev,
  Mod.\ Phys.\ Lett.\ A {\bf 26}, 1009 (2011).

\end{thebibliography}
\end{document}